\documentclass{article}

\usepackage{arxiv}

\usepackage[utf8]{inputenc} % allow utf-8 input
\usepackage[T1]{fontenc}    % use 8-bit T1 fonts
\usepackage{hyperref}       % hyperlinks
\usepackage{url}            % simple URL typesetting
\usepackage{booktabs}       % professional-quality tables
\usepackage{amsfonts}       % blackboard math symbols
\usepackage{nicefrac}       % compact symbols for 1/2, etc.
\usepackage{microtype}      % microtypography
\usepackage{lipsum}

\usepackage{amsfonts}
\usepackage{amsmath}
\usepackage{subcaption}
\usepackage{textcomp}
\usepackage{graphicx}
\usepackage{amssymb}
\usepackage[inkscapelatex=false]{svg}
\usepackage{mathtools}
\usepackage{newtxtext,newtxmath}
\usepackage{xcolor}
\usepackage{float}
\usepackage{algorithm}
\usepackage{algpseudocode}
\usepackage{stmaryrd}

\newtheorem{theorem}{Theorem}[section]

\newtheorem{corollary}[theorem]{Corollary}

\newtheorem{definition}[theorem]{Definition}
\newtheorem{lemma}[theorem]{Lemma}
\newtheorem{proof}[theorem]{Proof}
\newtheorem{remark}[theorem]{Remark}

\title{On Incremental Stability of Interconnected Switched Systems
}

\author{
 Bhabani Shankar Dey \\
  Centre for Cyber-Physical Systems\\
  IISc, Bengaluru, India\\
  \texttt{bhabanishankar440@gmail.com} \\
   \And
 Indra Narayan Kar \\
  Department of Electrical Engineering\\
  IIT Delhi, new Delhi, India\\
  \texttt{ink@ee.iitd.ac.in} \\
  \And
 Pushpak Jagtap \\
  Centre for Cyber-Physical Systems\\
  IISc, Bengaluru, India\\
  \texttt{pushpak@iisc.ac.in} \\
}

\begin{document}
\maketitle

\begin{abstract}
In this paper, the incremental stability of interconnected switched nonlinear systems is discussed. The nature of switching considered is state-dependent. The incremental stability of the switched interconnected system is a stronger property compared to the conventional notion of stability. Even if individual systems in the interconnected setting are stable, guaranteeing stability for the overall system is challenging. However, one of the important features of incremental stability is that the notion is preserved over interconnection. Here, leveraging the contraction-theoretic tools, we derive a set of sufficient conditions for the overall interconnection consisting of bimodal switched systems. To showcase the wider usability of our proposed results, we have also included the effect of external input, which leads to the study of incremental input-to-state stability ($\delta$-ISS). For the case of feedback interconnection, the small gain characterisation is presented for the overall system's $\delta$-ISS. Further, for a special case of feedback, i.e., cascade interconnection, the results are derived. The derived conditions are based on the matrix measure, making it computationally tractable and general. To make the results more general,  a generalised interconnection of bimodal switched systems is studied and corresponding sufficient condition for $\delta$-ISS are presented. Two numerical examples are demonstrated and supported with simulation results to verify the theoretical claims.
\end{abstract}


% \begin{highlights}
% \end{highlights}

%% Keywords



\section{Introduction}\label{sec1}
An interconnected system in control refers to a complex system made up of multiple subsystems that interact with each other. These subsystems include physical components, software systems, human operators, etc. They have numerous potential applications for network control systems, such as vehicle platooning, a series of biological networks, interconnected power grids, and transportation networks\cite{interconnect_1, interconnect_2, interconnect_3_swaroop}. In many applications, systems represent switching circuits, such as power converters, cascaded inverters, and switched mechanical systems \cite{7921574, pavlov2005convergent}. The behaviour of the interconnected system is determined by the interactions between the subsystems and how their individual characteristics affect the overall behaviour. In these systems, the interactions between the subsystems can significantly impact the system's overall performance, and any failures or disturbances in one subsystem can propagate to other subsystems and cause widespread disruptions. Due to the complexity of interconnected systems, the control, management, and stability of such systems require a vivid analysis. 

The methods of conducting stability analysis have evolved in multiple dimensions. One of the recent developments is the incremental stability analysis \cite{angeli2002Lyapunov}, \cite{zamani2011backstepping}, which is helpful in many applications where it is difficult to find the equilibrium state. To guarantee/analyse incremental behaviour, contraction theory has become an increasingly popular method. Unlike traditional Lyapunov stability analysis, which focuses on a system's attractor, i.e., an equilibrium point, contraction analysis examines the relative convergence of trajectories of a system. Various techniques exist for analysing the contractivity of smooth dynamical systems, such as those found in \cite{aminzare2014contraction, contractionmainpaper}. It has also been widely applied to consensus and synchronisation problems of networks \cite{delellis2010quad, Shiromoto}. For singularly perturbed nonlinear systems, high-gain observers based on contraction theory are designed in \cite{rayguru2022output}. The incremental stability has also been explored as a design methodology to guarantee certain performance measures \cite{jagtap2017backstepping, jagtap2020symbolic}. However, contrary to the applications of contraction theory to smooth systems, some work has been done in the switched system setting. For example, logarithmic norm-based contraction analysis for piecewise smooth systems has been formalised in \cite{di2014contraction}, \cite{fiore2017observer} and \cite{fiore2016contraction}. The concept of mode-dependent average dwell time for time-dependent switched systems was introduced in \cite{switched_contraction_bayu} based on contraction theory. In \cite{Incremental_impusle_NAHS}, the contraction of continuous-time dynamical systems with impulsive control is studied.

The study of interconnected systems has been of prime importance in the analysis and design of complex systems \cite{puspak}. Moreover, when the subsystems are switched systems, analysis becomes further complicated \cite{10008093}. A small-gain theorem is useful for analysing and designing non-switched interconnected nonlinear systems \cite{jiang1994small}, \cite{hill1977stability}. However, there has been limited work on applying this theorem to switched interconnected nonlinear systems \cite{switched_interconnected}. Recently, a Lyapunov theory-based small-gain theorem for time-dependent switched feedback interconnected systems was studied in \cite{smallgain_liberzon}. However, the convergence results are asymptotic with respect to the equilibrium point. Furthermore, the ISS small-gain theorem has been utilised to analyse the stability of hybrid integrator systems (plant is linear, controller is a hybrid system), which can be viewed as feedback interconnections of simpler subsystems \cite{van2023small}. Despite some attempts, the incremental stability of the interconnected system has been sparsely studied and still lacks generality \cite{swikir2019compositional}. If the interconnected systems are nonlinear switched systems, it adds to the challenges.
Preserving the usual notion of stability for the overall system is difficult for nonlinear switched systems. However, incremental stability using contraction has the strong property of preserving contractivity over interconnection. To the best of our knowledge, there has been no attempt to study the incremental stability of interconnected switched nonlinear systems.

Motivated by the above considerations, we derive sufficient conditions in terms of matrix measures to ensure incremental input-to-state stability of interconnected switched systems. In this paper, we mainly consider two classes of interconnections, namely the feedback connection and the cascade, to state the results. In both cases of interconnection, bimodal switched systems are considered. Finally, the $\delta$-ISS result corresponding to the general interconnection of $N$-bimodal switched subsystems is presented. The key highlights of the paper can be encapsulated as follows: 
\begin{itemize}
    \item Unlike the challenges faced in interconnected systems, our proposed framework leverages contraction theory's powerful property, which preserves incremental stability in interconnected nonlinear circuits/systems.
    \item Unlike results \cite{van2023small}, where the system considered is linear and feedback is switched, our proposition is valid for a more general interconnection. We consider that all interconnected subsystems are nonlinear switched systems based on state-dependent switching. This enhances the generality of the classes of systems under consideration.
    \item The use of contraction-theoretic tools help to derive sufficient conditions based on matrix measures to ensure the overall system's incremental ISS ($\delta$-ISS) with exponential convergence for both feedback and cascaded interconnections, whereas in many existing study of interconnected switched systems \cite{switched_interconnected, smallgain_liberzon}, the nature of convergence is asymptotic.
    \item We present a small gain-like condition to establish ($\delta$-ISS) of the overall system for the feedback system and later for the generalised interconnection. 
\end{itemize}
Section (\ref{Preliminary of Incremental Stability})  introduces the preliminaries related to the switched system in the presence of external and internal input. This section includes the definition of incremental ISS and preliminary existing results for the analysis of switched systems. Section (\ref{problem formulation}) discusses the basic problem setup with system descriptions. In section (\ref{main}), how contraction theory can be explored in analysing the ($\delta$-ISS) of switched interconnected systems is explained. While (\ref{generalised}) states the results for generalised interconnection. In (\ref{Results}), numerical examples are considered, and simulation results are presented to claim the propositions, followed by concluding remarks in the last section.

\textbf{Notations:} $\mathbb{R}$ is the set of real numbers. $\mathbb{R}_{\geq 0}$ is a set of non-negative real numbers.  A function $\alpha:\mathbb{R}_{\geq 0}\rightarrow \mathbb{R}_{\geq 0}$ is a $\mathcal{K}$ function if it strictly increases and $\alpha(0)=0$. $\gamma:\mathbb{R}_{\geq 0}\rightarrow \mathbb{R}_{\geq 0}$ is a $\mathcal{K_\infty}$ function if it is a $\mathcal{K}$ function and $lim_{s\rightarrow\infty} \gamma(s)=\infty$. $\beta:\mathbb{R}_{\geq 0}$ $\times$ $\mathbb{R}_{\geq 0}\rightarrow \mathbb{R}_{\geq 0}$ is said to be class $\mathcal{KL}$, if for all $ t\geq0$, $\beta(.,t)$ is a $\mathcal{K}$-function and for $\forall s\geq0$, $\beta(s,.)$ is non-increasing and it vanishes to zero at $t$ tending to infinity. A set $C\subseteq \mathbb{R}^n$ is defined to be forward invariant if for every initial condition $x(0)\in C$, $x(t)$ will lie in the set $C$ for all $t\geq0$. $||(.)||$ is the norm and $\mu(A)$ is the corresponding matrix measure. The gradient operator is denoted as $\nabla$. We require expressions to quantify the magnitude of the signals across specific time intervals. Since we need to assess the state value concerning the signal $\xi(.)$ over an interval encompassing $t$, we opt for the supremum norm to evaluate outputs. For a real-valued function $\xi:\mathbb{R}_{\geq 0}\rightarrow \mathbb{R}$ defined on the interval $[a,b]$. For each $0\leq a \leq b$, $||\xi||_{[a,b]}:= \sup\{||\xi(\tau)||: \tau\in[a,b]\}$. If $a=0$ refers to the initial time and $b=t$ refers to any time, the supremum can be written as $||\xi||_{t}$.

\section{Preliminaries and Background}\label{Preliminary of Incremental Stability}
\subsection{Piece-wise Smooth (PWS) Systems}
Consider a PWS system represented in the following form \cite{liberzon2003switching},
 \begin{equation}\label{pws_1}
\dot{\xi}=f_i(\xi, u_{int}, u_{ext}, t), \hspace{0.2 cm}\xi(t)\in S_i \subseteq{\mathbb{R}^{n}},
\end{equation}
where the state vector $\xi(t) \in \mathbb{X} \subseteq \mathbb{R}^n$, $u_{int}(t) \in \mathbb{U}_{int} \subseteq \mathbb{R}^{m_1}$ and $u_{ext}(t) \in \mathbb{U}_{ext} \subseteq \mathbb{R}^{m_2}$, where $u_{ext}(t)$ and $u_{int}(t)$ are the external and internal input/perturbation, respectively, and are Lebesgue measurable functions. Here, $i\in P$ represents that $i^{th}$ mode is active and $P$ is a finite index set defined as $P:=$$\{1, 2, 3,...,p\}$, $p$ is the number of switching subsystems. The vector field in \eqref{pws_1} lies in non-intersecting regions named $S_i$. $S_i$ are finite, disjoint, open sets such that $\cup_{i=1}^{p}\Bar{S}_i=\mathbb{X}\subseteq \mathbb{R}^n$ where $\Bar{S}_i=S_i\cup\partial S_i$. $\partial S_i$ corresponds to the boundary of region $S_i$. $\Bar{S}_i$ is a connected set and the intersection $\mathcal{M}:=\Bar{S}_i\cap\Bar{S}_j$ is the switching manifold (see Definition \ref{definition_manifold}). Vector fields $f_{i}(\xi,u_{int}, u_{ext}, t)$ are individually assumed to be smooth or Lipschitz at least locally, i.e., map $\xi\mapsto f_i(.)$ is continuously differentiable for all $i\in P$, and each $f_i$ is continuously extended on the boundary $\partial S_i$. The existence of a solution for \eqref{pws_1} is established in the sense of Filipov \cite{cortes2008discontinuous}, \cite{filippov2013differential}. The solutions are assumed to be well-posed.
\begin{definition}\label{definition_manifold}
  \cite{fiore2016contraction} Switching manifold $\mathcal{M}$ is defined as a zero set of a smooth function $H:\mathbb{R}^n\rightarrow \mathbb{R},$ i.e., 
   \begin{equation}
      \mathcal{M}:=\{x \in \mathbb{R}^n | H(x)=0\}, 
   \end{equation}
where $0\in\mathbb{R}$ is a regular value of $H$, $\forall x\in \mathcal{M}$, $\nabla H(x)=[\frac{\partial H(x)}{\partial x_1}...\frac{\partial H(x)}{\partial x_n}]\neq 0$. 
\end{definition}
If the system is bimodal, the switching manifold $\mathcal{M}$ divides the state-space into disjoint regions such as $\{S_1:=x\in\mathbb{R}^n:H(x)>0\}$ and $\{S_2:=x\in\mathbb{R}^n:H(x)<0\}$.

We use the notation $\xi_{xu}(t)$ to denote the solution of \eqref{pws_1} at time $t$ under the influence of the input signal $u$, starting from the initial state $x$.
\subsection{Incremental Input-to-State Stability}
Incremental stability characterises the time evolution of trajectories approaching each other. An efficient tool to establish incremental behaviour is \textit{contraction analysis}. In an incrementally stable system, if one of the trajectories is a steady-state solution, then all trajectories converge to the same. The switched system represented in \eqref{pws_1} is incrementally stable if it satisfies the relation in the following definition.  

\begin{definition}\label{definition_incremental}
\cite{angeli2002Lyapunov} The interconnected system in \eqref{pws_1} is said to be incremental input-to-state stable ($\delta-ISS$), if there exists a class $\mathcal{KL}$ function $\beta$, and  $\mathcal{K}_\infty$ functions $\gamma_{int}$, $\gamma_{ext}$ such that for any $t\in\mathbb{R}_{\geq 0}$,  any two initial conditions $x$, $x'\in \mathbb{X}$, any input signals $u_{int}, u'_{int}:[0,t] \rightarrow \mathbb{U}_{int}$ and $u_{ext}$, $u'_{ext}:[0,t] \rightarrow \mathbb{U}_{ext}$, the following condition is satisfied:
% \textcolor{red}{ Also $u^1$, $u^2$ are not defined!},  
\begin{align}\label{kl_incremental_input_def} 
    ||\xi_{xu}(t)-\xi_{x'u'}(t)||
    \leq\beta(||x-x'||, \hspace{0.1 cm} t)+\gamma_{int}(||u_{int}-u'_{int}||_t)+\gamma_{ext}(||u_{ext}-u'_{ext}||_t).
\end{align}  
\end{definition} 
The notation used in the subscript in the left-hand side of \eqref{kl_incremental_input_def} is defined as $u=(u_{int}, u_{ext})$ and $u'=(u'_{int}, u'_{ext})$. It can be observed that in the absence of external/internal input ($u(t), u'(t)\equiv 0$), $\delta-ISS$ in \eqref{kl_incremental_input_def} reduces to incremental Global Asymptotic Stability ($\delta-GAS$). Furthermore, if the function $\beta$ in the above definition takes an exponentially decaying form, i.e., $||x-x'||e^{-ct}$, for $c>0$, then the notion of stability is referred to as incremental Global Exponential Stability ($\delta$-GES).\\

In the purview of incremental stability, contractivity of system dynamics is equivalent to the exponential stability of its corresponding variational dynamics \cite{contractionmainpaper}. Therefore, the problem is reduced to studying the exponential stability of variational systems, which will be defined appropriately in the subsequent section. To prove the contractivity of the vector field, the dynamics are assumed to be continuously differentiable, which necessitates that the derivatives are continuous. Unfortunately, the currently focused problem violates this assumption because of the switching, and hence, the problem demands to be solved in a different setup.  As we have already discussed, the smoothness assumption for the usual contraction analysis is violated, and the dynamics undergo a regularisation process. Regularisation is a transformation adopted for the analysis of non-smooth systems with the help of transition functions (refer to \cite{johansson1999regularization,mynttinen2015smoothing} for more details). All convergence analyses are performed using the concept of matrix measure. First, we present some definitions and lemmas necessary to state the main results. These results correspond to the system represented in \eqref{pws_1}, but in the absence of input, i.e., $u_{int}=u_{ext}=0$.
\begin{definition}
   \cite{fiore2016contraction} Given a Piece-wise Smooth Continuous (PWSC) transition function $\zeta: \mathbb{R}\rightarrow [-1,1]$ defined as
   \begin{equation}\label{transition_function}
        \zeta(r)=\begin{cases}
       1 & \text{for $r \geq 1$},\\
      \in(-1, 1) & \text{for $r\in(-1, 1)$},\\
      -1 & \text{for $r \leq -1$},
    \end{cases}
    \end{equation}
    and $\frac{d}{dr}\zeta>0$ for $r\in(-1, 1)$.
\end{definition}
\begin{definition}
  \cite{fiore2016contraction} The $\zeta$-regularisation of two vector fields $f_1$ and $f_2$ of PWS system, for parameter $\epsilon>0$ is given by 
   \begin{equation}\label{convex_combi}
    f_\epsilon(x)=\Psi(x)f_1+\Gamma(x)f_2,
\end{equation}
where
\begin{equation*}
     \Psi(x)= \frac{1}{2}\Bigg[1+\zeta\Big(\frac{H(x)}{\epsilon}\Big)\Bigg]\text{ and
}     \Gamma(x)= \frac{1}{2}\Bigg[1-\zeta\Big(\frac{H(x)}{\epsilon}\Big)\Bigg].
\end{equation*} 
Functions $\Psi(x)\in [0, 1]$, $\Gamma(x)\in [0, 1]$ such that $\Psi(x)+\Gamma(x)=1$, $x\in\mathbb{X}$ and $H(x)$ is a smooth function as mentioned in Definition \ref{definition_manifold}.
\end{definition}

\begin{definition}\label{matrix_measure}
    \cite{vidyasagar1978matrix} The matrix measure or logarithmic norm $\mu$ of a square matrix $A$ is defined as 
\begin{equation}
    \mu(A)=\lim_{h \rightarrow 0^+ }\sup \frac{||(I_n+hA)||-1}{h}.
\end{equation}
\end{definition}
Depending on the induced norm, the computation of the matrix measure changes. Certain important properties of matrix measure that will be used in this paper are mentioned here. For square matrices $A$, $B$, and for all $\lambda \in [0,1]$, \begin{align}\label{property_mu}
    & P1 (\textit{Subadditivity}): \hspace{0.1 cm}\mu(A+B)\leq \mu(A)+\mu(B), \hspace{0.1 cm}\nonumber\\ 
     & P2 (\textit{Convexity}): \hspace{0.1 cm}\mu(\lambda A+ (1-\lambda) B)\leq \lambda \mu(A)+(1-\lambda)\mu(B).
 \end{align}
The following lemma presents sufficient conditions for a PWS system's incremental stability, consisting of two modes.
\begin{lemma}\label{main_lemma}
\cite{fiore2016contraction} Let $C\subseteq \mathbb{R}^n$ be some forward invariant set. If there exist some positive constants $c_1$, $c_2>0$, and certain norm defined in $C$, and corresponding induced matrix measure denoted as $\mu$ (as defined in Definition \ref{matrix_measure}), then a bimodal switched system (i.e., $P=\{1,2\}$ in \eqref{pws_1}) with $u=(u_{int}, u_{ext})=0$ is said to be incrementally exponentially stable in $C$ with a convergence rate $c=\min(c_1, c_2)$, (i.e., for all $t\geq 0$, $||\delta \xi(t)||\leq ke^{-ct}||\delta \xi(0)||$),
if for all $t\geq 0$ the following conditions hold. 
\begin{equation}
    \label{cond1}
        \mu \Big(\frac{\partial f_1(\xi, t)}{\partial \xi}\Big)\leq -c_1, \hspace{0.1 cm}\forall \xi(t)\in\Bar{S}_1,
    \end{equation}
    \begin{equation}
    \label{cond2}
        \mu \Big(\frac{\partial f_2(\xi, t)}{\partial \xi}\Big)\leq -c_2, \hspace{0.1 cm}\forall \xi(t)\in\Bar{S}_2,
    \end{equation}
     \begin{equation}
     \label{cond3}
        \mu [(f_1(\xi, t)-f_2(\xi, t))\nabla H(\xi(t))]=0, \hspace{0.1 cm}\forall \xi(t)\in \mathcal{M}.
\end{equation}
\end{lemma}
Here, $\delta \xi$ is the virtual displacement allowed in the solution of \eqref{pws_1} in the tangent direction. The detailed proof of the Lemma \ref{main} can be referred to in \cite[Theorem-6]{fiore2016contraction}. The following lemma details the connection between the incremental stability of the regularised system and the original PWS system.
% \textcolor{red}{what happens to inputs?}
\begin{lemma}\label{regular_lemma}
In a forward invariant set $C\subseteq \mathbb{R}^n$, if there exists some $\epsilon >0$, the regularized system is exponentially incrementally stable with a convergence rate $c>0$, any two solutions of the original system $\xi_{x}(t)$ and $\xi_{x'}(t)$ for any initial states $x, x' \in C$ converge towards each other in the limiting behavior, i.e., $\epsilon \rightarrow 0$. 
\end{lemma}
The proof of this lemma can be found in \cite{fiore2016contraction}.
\section{System Description and Problem Formulation}\label{problem formulation}
\subsection{Interconnected System}
\begin{figure}[!ht]
		\centering	\includegraphics[width=0.5\textwidth]{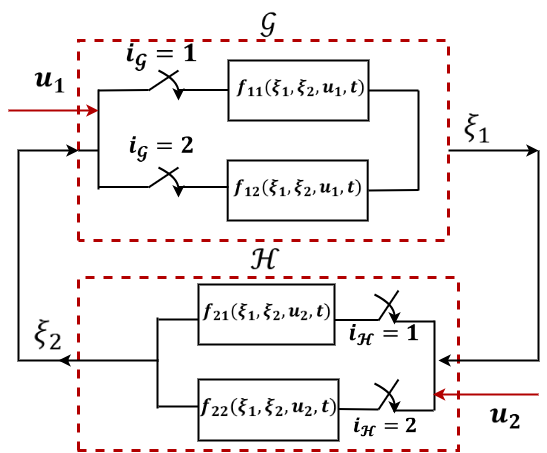}
		\caption{Feedback interconnection of two switched nonlinear systems} 
  \label{schematic_diagram_2}
\end{figure}

There are numerous kinds of interconnected systems, including parallel, feedback, cascade, and others, based on how they are connected to each other. This work primarily examines a feedback-type interconnection, establishing fundamental results. It then explores a cascade connection, which is being treated as a special case of feedback interconnection, before extending the analysis to a general interconnection. Corresponding theoretical results are proposed to provide a comprehensive understanding of system behaviour under interconnection. First, we consider a feedback interconnection, which is an important class of interconnected systems having significant applications in control system design, as shown in Fig.~\ref{schematic_diagram_2}. It can be inferred from the figure that two subsystems $\mathcal{G}$ and $\mathcal{H}$ are interconnected and are individually switched nonlinear systems. To see the effect of external input/perturbations in interconnection, $u_1$ and $u_2$ are considered. Moreover, the output(state in this case) of $\mathcal{G}$ acts as an internal input to $\mathcal{H}$ and vice versa. The feedback interconnection of switched non-linear systems perturbed by external and internal inputs can be represented as follows:
\begin{gather}\label{interconnected_feedback}
    \begin{split}
\mathcal{G}:\dot{\xi_1}&=f_{1i_\mathcal{G}}(\xi_1, \xi_2, u_1, t),\\
\mathcal{H}: \dot{\xi_2}&=f_{2i_{\mathcal{H}}}(\xi_1, \xi_2, u_2, t),
    \end{split}
\end{gather}
where $i_{\mathcal{G}}$ and $i_{\mathcal{H}}$ are the switching signals corresponding to two systems $\mathcal{G}$ and $\mathcal{H}$, respectively. Compared to the general representation in \eqref{pws_1}, for the $\mathcal{G}$ (or $\mathcal{H}$) system, $\xi_2$ (or $\xi_1$) acts as an internal input, i.e., $u_{int}(t)$ and $u_1(t)$ (or $u_2(t)$) acts as an external input, i.e., $u_{ext}(t)$. In this paper, each switched system in the interconnection is assumed to be bimodal, i.e., each switched system has two subsystems. Here, switching is used based on state-dependent conditions. The overall interconnection in compact form can be
\begin{equation}\label{feedback_interconnected}
  \mathcal{G} :
    \begin{cases}
       \dot{\xi_1}=f_{11}(\xi_1, \xi_2, u_1, t) & \text{for $i_{\mathcal{G}}=1$},\\
      \dot{\xi_1}=f_{12}(\xi_1, \xi_2, u_1, t) & \text{for $i_{\mathcal{G}}=2$}.\\
    \end{cases}   
\end{equation}
\begin{equation}\label{feedback_interconnected_1}
 \mathcal{H} : 
    \begin{cases}
       \dot{\xi_2}=f_{21}(\xi_1, \xi_2, u_2, t) & \text{for $i_{\mathcal{H}}=1$},\\
      \dot{\xi_2}=f_{22}(\xi_1, \xi_2, u_2, t) & \text{for $i_{\mathcal{H}}=2$}.\\
    \end{cases}   
\end{equation}
\subsection{Problem Statement}
The main objective is divided into two parts. The first task is to study the incremental stability of the individual nonlinear switched systems. The second objective is to derive sufficient conditions to ensure the  ($\delta$-ISS) of the overall interconnected system in the presence of external input. 
 %\vspace{-0.6 cm}
\section{Main Result}\label{main}
\begin{figure}[!ht]
		\centering	\includegraphics[width=0.8\textwidth]{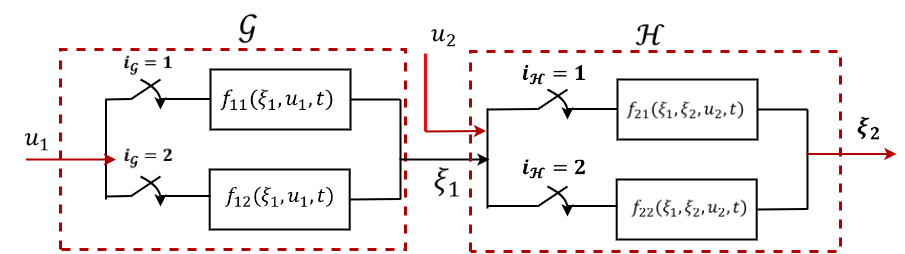}
		\caption{Cascade connection of two switched nonlinear systems}
  \label{schematic_diagram}%\vspace{-0.6cm}
\end{figure}

As discussed, we must see the variational system's ISS to prove $\delta$-ISS of the overall switched interconnection represented in \eqref{interconnected_feedback}. Defining a variational system for non-smooth systems requires regularisation of the vector fields. Therefore, the regularised vector fields for both systems are represented as follows:
\begin{equation}\label{reg_3}
    f_{\epsilon\mathcal{G}}(\xi_1, t)=\psi(\xi_1)f_{11}(\xi_1, \xi_2, u_1, t)+\Gamma(\xi_1) f_{12}(\xi_1, \xi_2, u_1, t),
\end{equation}
\begin{equation}\label{reg_4}
    f_{\epsilon\mathcal{H}}(\xi_2,t)=\psi(\xi_2)f_{21}(\xi_1, \xi_2, u_2, t)+\Gamma(\xi_2) f_{22}(\xi_1, \xi_2, u_2, t).
\end{equation}
where $f_{\epsilon\mathcal{G}}$, $f_{\epsilon\mathcal{H}}$ are regularised vector fields corresponding to $\mathcal{G}$ and $\mathcal{H}$ systems, respectively. Since, from lemma \ref{regular_lemma}, it is evident that the contractivity of the regularised system is equivalent to that of the PWS when $\epsilon\rightarrow0$, we investigate the contraction of the regularised vector fields. The variational dynamics for both the vector fields in \eqref{reg_3} and \eqref{reg_4} are given as
\begin{equation}\label{variation_feedback_1}
   \dot {\delta\xi_1}=\frac{\partial f_{\epsilon\mathcal{G}}(\xi_1,t)}{\partial \xi_1}\delta \xi_1+\frac{\partial f_{\epsilon\mathcal{G}}(\xi_1,t)}{\partial \xi_2}\delta \xi_2+\frac{\partial f_{\epsilon\mathcal{G}}(\xi_1,t)}{\partial u_1}\delta u_1,
\end{equation}
\begin{equation}\label{variation_feedback_2}
   \dot {\delta\xi_2}=\frac{\partial f_{\epsilon\mathcal{H}}(\xi_2,t)}{\partial \xi_2}\delta \xi_2+\frac{\partial f_{\epsilon\mathcal{H}}(\xi_2,t)}{\partial \xi_1}\delta \xi_1+\frac{\partial f_{\epsilon\mathcal{H}}(\xi_2,t)}{\partial u_2}\delta u_2.
\end{equation}
Further simplifying and arranging the similar terms together, time differentiation of the variational state of $\mathcal{G}$-system, 
\begin{align}\label{var_5_1}    \dot {\delta\xi_1}&=\underbrace{\Bigg(\psi(\xi_1)\frac{\partial f_{11}(\xi_1, \xi_2, u_1, t)}{\partial \xi_1}+\Gamma(\xi_1)\frac{\partial f_{12}(\xi_1, \xi_2, u_1, t)}{\partial \xi_1}\Bigg)}_\text{$F_1$}\delta \xi_1\nonumber\\
    &+ \underbrace{\Bigg(\psi(\xi_1)\underbrace{\frac{\partial f_{11}(\xi_1, \xi_2, u_1, t)}{\partial \xi_2}}_\text{$F_{11}$}+\Gamma(\xi_1)\underbrace{\frac{\partial f_{12}(\xi_1, \xi_2, u_1, t)}{\partial \xi_2}}_\text{$F_{12}$}\Bigg)}_\text{$F_2$}\delta \xi_2\nonumber\\
    &+\underbrace{\Bigg(f_{11}(\xi_1, \xi_2, u_1, t)\frac{\partial \psi(\xi_1)}{\partial \xi_1}+f_{12}(\xi_1, \xi_2, u_1, t)\frac{\partial \Gamma(\xi_1)}{\partial \xi_1}\Bigg)}_\text{$F_3$}\delta \xi_1\nonumber\\
    &+\underbrace{\Bigg(\psi(\xi_1)\frac{\partial f_{11}(\xi_1, \xi_2, u_1, t)}{\partial u_1}+\Gamma(\xi_1)\frac{\partial f_{12}(\xi_1, \xi_2, u_1, t)}{\partial u_1}\Bigg)}_\text{$F_4$}\delta u_1.
\end{align}
Similarly, for $\mathcal{H}$-system,
 \begin{align}\label{var_5_2}
    %\begin{split}
    \dot {\delta\xi_2}&=\underbrace{\Bigg(\psi(\xi_2)\frac{\partial f_{21}(\xi_1, \xi_2, u_2, t)}{\partial \xi_2}+\Gamma(\xi_2)\frac{\partial f_{22}(\xi_1, \xi_2, u_2, t)}{\partial \xi_2}\Bigg)}_\text{$G_1$}\delta \xi_2\nonumber\\
    &+  \underbrace{\Bigg(\psi(\xi_2)\underbrace{\frac{\partial f_{21}(\xi_1, \xi_2, u_2, t)}{\partial \xi_1}}_\text{$G_{21}$}+\Gamma(\xi_2)\underbrace{\frac{\partial f_{22}(\xi_1, \xi_2, u_2, t)}{\partial \xi_1}}_\text{$G_{22}$}\Bigg)}_\text{$G_2$}\delta \xi_1\nonumber\\
    &+\underbrace{\Bigg(f_{21}(\xi_1, \xi_2, u_2, t)\frac{\partial \psi(\xi_2)}{\partial \xi_2}+f_{22}(\xi_1, \xi_2, u_2, t)\frac{\partial \Gamma(\xi_2)}{\partial \xi_2}\Bigg)}_\text{$G_3$}\delta \xi_2,\nonumber\\
    &+\underbrace{\Bigg(\psi(\xi_1)\frac{\partial f_{21}(\xi_1, \xi_2, u_2, t)}{\partial u_2}+\Gamma(\xi_1)\frac{\partial f_{22}(\xi_1, \xi_2, u_2, t)}{\partial u_2}\Bigg)}_\text{$G_4$}\delta u_2.
   % \end{split}
\end{align}
In compact form, one can rewrite \eqref{var_5_1} and \eqref{var_5_2} as
\begin{equation}\label{variational_sys_1}
    \dot {\delta\xi_1}=(F_1+F_3)\delta \xi_1+F_2 \delta \xi_2+F_4\delta u_1,
\end{equation}
\begin{equation}\label{variational_sys_2}
    \dot {\delta\xi_2}=(G_1+G_3)\delta \xi_2+G_2 \delta \xi_1+G_4\delta u_2.
\end{equation}
First, we discuss the ISS of the individual variational systems in \eqref{variational_sys_1} and \eqref{variational_sys_2}. The results are stated in the form of Theorem \ref{theorem_3}.
Assume that $j$ is the counter that refers to the number of interconnected switched systems, and $i$ refers to the number of modes in each switched system. In this paper, first, we consider both $i$, $j\in\{1, 2\}$. Later, we state the result for a general interconnected switched system. 
\begin{theorem}\label{theorem_3}
Consider the variational systems in \eqref{variational_sys_1} and \eqref{variational_sys_2}. Let $C\subseteq \mathbb{R}$$^n$ be a forward invariant set. If there exist some norm in ${C}$ and associated matrix measure $\mu$ such that for some positive constants $c_i$, $d_i$, $\forall i\in\{1, 2\}$ such that,
\begin{equation}\label{interconneted_conditions}
\begin{split}
    1) &\hspace{0.1 cm}  \mu\Bigg(\frac{\partial f_{ji}(\xi_1, \xi_2, t)}{\partial \xi_j}\Bigg)\leq -c_i,\hspace{0.2 cm}\forall \xi_j \in \Bar{S}_{ji}, \hspace{0.1 cm}\text{for}\hspace{0.1 cm} j=1, \\
    2) &\hspace{0.1 cm} \mu\Bigg(\frac{\partial f_{ji}(\xi_1, \xi_2, t)}{\partial \xi_j}\Bigg)\leq -d_i, \hspace{0.2 cm}\forall \xi_i \in \Bar{S}_{ji}, \hspace{0.1 cm}\text{for}\hspace{0.1 cm} j=2,\\
    3) &  \hspace{0.1 cm} \mu[(f_{j1}-f_{j2})\nabla H(\xi_j)]=0, \hspace{0.2 cm}\forall \xi_j \in{\mathcal{M}_j}\hspace{0.1 cm} \forall j\in\{1,2\}.
\end{split}
\end{equation}
where $\mathcal{M}_j$ refers to the switching manifolds for each switched system in the interconnection.
Then, \eqref{variational_sys_1} is ISS with respect to $(\delta\xi_2, \delta u_1)$ with gain $\gamma_{int\mathcal{G}}$ from $\delta\xi_2$ to $\delta\xi_1$, i.e.,
\begin{equation}\label{ISS_1}   ||\delta\xi_1(t)||\leq\beta_1(||\delta\xi_1(0)||, t)+ \gamma_{int\mathcal{G}}(||\delta\xi_2||_t)+\gamma_{ext\mathcal{G}}(||\delta u_1||_t),
\end{equation}
and \eqref{variational_sys_2} is ISS with respect to $(\delta\xi_1, \delta u_2)$ with gain $\gamma_{int\mathcal{H}}$ from $\delta\xi_1$ to $\delta\xi_2$, i.e.,
\begin{equation}\label{ISS_2}    ||\delta\xi_2(t)||\leq\beta_2(||\delta\xi_2(0)||, t)+ \gamma_{int\mathcal{H}}(||\delta\xi_1||_t)+\gamma_{ext\mathcal{H}}(||\delta u_2||_t),
\end{equation}
for some $\beta_1$, $\beta_2$ $\in\mathcal{KL}$, $\gamma_{int\mathcal{G}}$, $\gamma_{ext\mathcal{G}}$, $\gamma_{int\mathcal{H}}$, $\gamma_{ext\mathcal{H}}$ $\in \mathcal{K}_\infty$.
\end{theorem}
\begin{proof}
To upper bound the evolution of the variational state in \eqref{variational_sys_1}, we apply the upper Dini derivative to $||\delta \xi_1(t)||$,
\begin{equation}\label{measure}
    \begin{split}
        D^{+}||{\delta \xi_1(t)}||=
    \lim_{h \rightarrow 0^+ }\sup \frac{||\delta \xi_1(t+h)||-||\delta \xi_1(t)||}{h},\\
    \leq
    \lim_{h \rightarrow 0^+ }\sup \frac{||\delta \xi_1(t)+h\dot {\delta \xi_1}||-||\delta \xi_1(t)||}{h}.     
    \end{split}
\end{equation}
Replacing the differential dynamics given in \eqref{variational_sys_1} in the above expression, the Dini derivative $D^{+}||{\delta\xi_1(t)}||$ can be written as
{\small\begin{equation}
    \begin{split}
        &D^{+}||{\delta\xi_1(t)}||\\&\leq  \lim_{h \rightarrow 0^+ } \frac{||\delta \xi_1+h[(F_1+F_3)\delta \xi_1+F_2\delta \xi_2+F_4\delta u_1]||-||\delta \xi_1(t)||}{h}\\
       &= \lim_{h \rightarrow 0^+ }\frac{||(I_n+h(F_1+F_3))\delta \xi_1+hF_2\delta \xi_2+hF_4\delta u_1||-||\delta \xi_1||}{h}\\
       &\leq \lim_{h \rightarrow 0^+ } \frac{(||I_n+h(F_1+F_3)||-1)||\delta \xi_1||+||hF_2\delta \xi_2||+||hF_4\delta u_1||}{h}\\
       &\leq \lim_{h \rightarrow 0^+ } \frac{(||I_n+h(F_1+F_3)||-1)}{h}||\delta \xi_1(t)||+||F_2||||\delta \xi_2(t)||+||F_4||||\delta u_1(t)||.
    \end{split}
\end{equation}}
Using the matrix measure in Definition \ref{matrix_measure}, the above expression can be rewritten as
\begin{equation}\label{differential_1}
    \begin{split}
       D^{+}||{\delta\xi_1(t)}|| \leq 
       \mu(F_1+F_3)||\delta \xi_1||+||F_2||||\delta \xi_2||+||F_4||||\delta u_1(t)||.
    \end{split}
\end{equation}
Similarly, for the $\mathcal{H}$-system,
\begin{equation}\label{differential_2}
    \begin{split}
       D^{+}||{\delta\xi_2}(t)||\leq \mu(G_1+G_3)||\delta \xi_2||+||G_2||||\delta \xi_1||+||G_4||||\delta u_2(t)||.
    \end{split}
\end{equation}
If we closely look at the $F_3$ term in \eqref{var_5_1} and evaluate the individual terms using the definition in \eqref{transition_function} and \eqref{convex_combi}, 
\begin{equation}\label{jacobian_1}
\begin{split}
    \frac{\partial \Psi(\xi_1)}{\partial \xi_1}=\frac{1}{2}\frac{\partial \zeta}{\partial r}\Bigg(\frac{H(\xi_1)}{\epsilon}\Bigg)\frac{\partial}{\partial \zeta}\Bigg[\frac{H(\xi_1)}{\epsilon}\Bigg]
    =\chi(\xi_1)\nabla H(\xi),
\end{split}   
\end{equation}
where $\chi(\xi_1)= \frac{1}{2\epsilon}\frac{\partial \zeta}{\partial r}\Bigg(\frac{H(\xi_1)}{\epsilon}\Bigg)$, $r=\frac{H(\xi_1)}{\epsilon}$.\\
Similarly,
\begin{equation*}\label{anti_parallel}
    \frac{\partial \Gamma(\xi_1)}{\partial \xi_1}=-\frac{\partial \Psi(\xi_1)}{\partial \xi_1}.
\end{equation*}
Putting them together and rewriting,
\begin{equation}
    F_3=\chi(.)[(f_{11}-f_{12})\nabla H(\xi_1)].
\end{equation}
Similar to the approach in \cite{fiore2016contraction}, to restrict the trajectories from diverging while switching, the matrix measure $[(f_{11}-f_{12})\nabla H(\xi_1)]$ is forced to zero. Hence, the matrix measure of $F_3$ becomes
\begin{equation}
    \mu(F_3)=\chi(\xi_1)\mu[(f_{11}-f_{12})\nabla H(\xi_1)]=0.
\end{equation}
Similarly, the following holds true for the $G_3$ term in \eqref{var_5_2}.
\begin{equation}
    \mu(G_3)=\chi(\xi_2)\mu[(f_{21}-f_{22})\nabla H(\xi_2)]=0.
\end{equation}
This proves the condition-3 of Theorem \ref{theorem_3} for $j\in\{1, 2\}$.
Therefore \eqref{differential_1} and \eqref{differential_2} can be written as 
\begin{equation}\label{dynamics_1}
    \begin{split}
       D^{+}||{\delta\xi_1(t)}||\leq 
       \mu(F_1)||\delta \xi_1(t)||+||F_2||||\delta \xi_2(t)||+||F_4||||\delta u_1(t)||.
    \end{split}
\end{equation}
\begin{equation}\label{dynamics_2}
    \begin{split}
       D^{+}||{\delta\xi_2}(t)||\leq \mu(G_1)||\delta \xi_2(t)||+||G_2||||\delta \xi_1(t)||+||G_4||||\delta u_2(t)||.
    \end{split}
\end{equation}
By simple integration of \eqref{dynamics_1}, the corresponding trajectory is written as
\begin{equation}\label{gronwal_1}
    \begin{split}
       ||\delta\xi_1(t)|| \leq e^{\mu(F_1)t}||\delta \xi_1(0)||
       &+\int_0^t e^{\mu(F_1)(t-\tau)}||F_2||||\delta \xi_2(\tau)||+\int_0^t e^{\mu(F_1)(t-\tau)}||F_4||||\delta u_1(\tau)||)d\tau.
    \end{split}
\end{equation}
On further simplification, $||\delta\xi_1(t)||$ 
\begin{align}
\label{gronwal_2}
\leq e^{\mu(F_1)t}||\delta \xi_1(0)||+e^{\mu(F_1)t}(||F_2||)\int_0^t e^{\mu(F_1)(-\tau)}(||\delta \xi_2(\tau)||\nonumber
+||F_4||||\delta u_1(\tau)||)d\tau,\nonumber
\end{align}
\begin{align*}
 \leq e^{\mu(F_1)t}||\delta \xi_1(0)||\nonumber
+e^{\mu(F_1)t}||F_2||\sup_{0\leq \tau \leq t}||\delta \xi_2(\tau)||\int_0^t e^{\mu(F_1)(-\tau)}d\tau,\nonumber   
\end{align*}
\begin{align}
  \leq e^{\mu(F_1)t}||\delta \xi_1(0)||+\gamma_1\sup_{0\leq \tau \leq t}||\delta \xi_2(\tau)||+\gamma_2\sup_{0\leq \tau \leq t}||\delta u_1(\tau)||.  
\end{align}
Similarly, for the second variational dynamics \eqref{dynamics_2}, 
%\vspace{-0.8 cm}
\begin{align}\label{gronwal_3}
& \nonumber\\
       &||\delta\xi_2(t)||\leq e^{\mu(G_1)t}||\delta \xi_2(0)||+\gamma_3\sup_{0\leq \tau \leq t}||\delta \xi_1(\tau)||+\gamma_4\sup_{0\leq \tau \leq t}||\delta u_2(\tau)||,
\end{align}
where $\gamma_1=\frac{||F_2||}{|\mu(F_1)|}$, $\gamma_2=\frac{||F_4||}{|\mu(F_1)|}$, $\gamma_3=\frac{||G_2||}{|\mu(G_1)|}$, $\gamma_4=\frac{||G_4||}{|\mu(G_1)|}$. \\
Using the representation of supremum in the notation section and the arguments thereafter, \eqref{gronwal_2} and \eqref{gronwal_3} can be written as
\begin{equation}  
\begin{split}\label{gronwal_4}
       ||\delta\xi_1(t)|| \leq e^{\mu(F_1)t}||\delta \xi_1(0)||+\gamma_1||\delta \xi_2(\tau)||_t+\gamma_2||\delta u_1(\tau)||_t.
\end{split}
\end{equation}
\begin{equation}\label{gronwal_5}
    \begin{split}
       ||\delta\xi_2(t)|| \leq e^{\mu(G_1)t}||\delta \xi_2(0)||+\gamma_3||\delta \xi_1(\tau)||_t+\gamma_4||\delta u_2(\tau)||_t.
    \end{split}
\end{equation}
In this case, $\gamma_{l}(r)$'s are linear in $r$ i.e., $\gamma_{l}(r)=\gamma_{l}r$ for $l\in\{1,2,3,4\}$, and they qualify as $\mathcal{K_\infty}$ functions as defined in the notations section. If we invoke condition-1, for $j=1$, $\forall i\in{1, 2}$, condition-2 for $j=2$, $\forall i\in{1, 2}$ in Theorem \ref{theorem_3}, by using the property of matrix measure as mentioned in \eqref{property_mu} and the regularised Jacobian in \eqref{var_5_1}, \eqref{var_5_2}, $\mu(F_1)$ and $\mu(G_1)$ can be negative. In such case, the functions $e^{\mu(F_1)t}$ and $e^{\mu(G_1)t}$ are exponentially decaying. As exponentially decaying functions are $\mathcal{KL}$, we represent $\|\delta \xi_1(0)\|e^{\mu(F_1)t}=\beta_1$ and $\|\delta \xi_2(0)\|e^{\mu(G_1)t}=\beta_2$, where $\beta_1, \beta_2\in\mathcal{KL}$. Hence, it can be observed that \eqref{gronwal_4}, and \eqref{gronwal_5} are exact relations for $\delta$-ISS of individual systems as given with \eqref{ISS_1} and \eqref{ISS_2} respectively. Where, $\gamma_{int\mathcal{G}}=\gamma_1$, $\gamma_{ext\mathcal{G}}=\gamma_2$, $\gamma_{int\mathcal{H}}=\gamma_3$, and $\gamma_{ext\mathcal{H}}=\gamma_4$. This completes the proof of Theorem \ref{theorem_3}.
\end{proof}
\begin{remark}
  Theorem \ref{theorem_3} is primarily aimed at establishing $\delta-$ISS of individual subsystems. It is important to note that \eqref{ISS_1}, \eqref{ISS_2} represent the $\delta$-ISS relations if the sufficient conditions in Theorem \ref{theorem_3} are satisfied. However, these relations are based on the evolution of infinitesimal differential lengths $\delta \xi_1$ and $\delta \xi_2$. This is equivalent to studying the time evolution of the distance between two trajectories as in definition \ref{definition_incremental}. This equivalence between the two approaches to show incremental stability is well established in the literature and can be seen in \cite{kawano2024incremental}.
\end{remark}
Once we ensure individual subsystems in an interconnection are $
\delta$-ISS,  we need to comment on the $\delta$-ISS of the overall interconnection. The arguments are similar to those of small gain conditions in the literature \cite{jiang1994small}, \cite{liberzon2006stability}. The result is stated in the following theorem.
\begin{theorem}\label{final_theorem}
    Consider that for the overall interconnected system, all conditions in Theorem-\ref{theorem_3} are satisfied such that,
    \begin{itemize}
        \item $\mathcal{G}$ is $\delta$-ISS with respect to $(\xi_2, u_1)$ with gain $\gamma_1$ from $\xi_2$ to $\xi_1$,
        \item $\mathcal{H}$ is $\delta$-ISS with respect to $(\xi_1, u_2)$ with gain $\gamma_3$ from $\xi_1$ to $\xi_2$,         
    \end{itemize}
and if there exists a function $\rho\in \mathcal{K}_\infty$ such that
\begin{equation}\label{small_gain_condition}
(id+\rho)\circ\gamma_1\circ(id+\rho)\circ\gamma_3(r)<r, \hspace{0.1cm} \forall r\geq0.
\end{equation}
Then, the overall feedback interconnection is $\delta$-ISS with respect to the input $(u_1, u_2)$. Here, $id$ is an identity function.
\end{theorem}
\begin{remark}
If the gain functions $\gamma_l(r)$ are a linear functions, i.e., $\gamma_l(r)$=$\gamma_{l}r$, for $l\in\{1, 2, 3, 4\}$, then the small gain condition in \eqref{small_gain_condition} reduces to $\gamma_1\gamma_3<1$. This condition resembles the standard small-gain results available in the literature. The proof of Theorem \ref{final_theorem} is readily inspired from the standard small gain theorem for nonlinear interconnected systems \cite{jiang1994small}. In \cite{jiang1994small}(Theorem 2.1), it uses the input-output practical stability(IOpS) and the property of unboundedness observability(UO), to comment on the IOpS of the interconnection. The resulting condition is a generalised small-gain relation as given in \eqref{small_gain_condition}. If the output is the same as the state, and the practical stability margin is assumed to be zero, then the concept of IOpS reduces to ISS. In our result, one of the contrasting factors is that instead of individual subsystems being simple nonlinear systems, we consider nonlinear state-dependent switched systems. One more standout difference is that instead of looking at the evolution of the state(with respect to the equilibrium), we are interested in analysing the incremental state evolution. Therefore, the sketch of the proof remains the same apart from considering the incremental state. The necessity of ensuring ISS for individual subsystems modifies to satisfy $\delta$-ISS, which has already been addressed in Theorem \ref{theorem_3}.
\end{remark}
\textbf{Special Cases:} It is interesting to highlight two special cases of Theorem \ref{final_theorem} explicitly.\\
\emph{Case-1:} In the first case, if there is no external input signal, i.e., $u_1=u_2\equiv0$, we conclude that the overall interconnected system is globally incrementally stable ($\delta$-GAS).\\
\emph{Case-2:} In the second case, Theorem \ref{final_theorem} also covers the cascaded connection. This is essentially a special case of the feedback as in the schematics shown in Fig.~\ref{schematic_diagram_2} if the link of the output of $\mathcal{H}$-system, which acts as an input to the $\mathcal{G}$-system gets broken, we achieve a cascade connection as shown in Fig.~\ref{schematic_diagram}. Then 
$\mathcal{G}$ is $\delta$-ISS with respect to $(u_1)$, and the small gain condition automatically holds. Hence, the overall cascaded interconnection is $\delta$-ISS. Therefore, in this article, we don't explicitly attempt to derive the results for the cascade connection.
%\vspace{-0.8 cm}
\section{Generalised Interconnection of Bimodal Switched Systems}\label{generalised}
\begin{figure}[!ht]
		\centering	\includegraphics[width=0.6\textwidth]{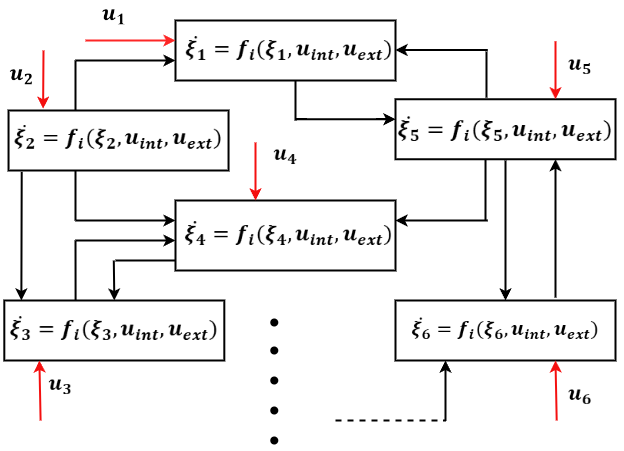}
		\caption{General interconnection of switched nonlinear systems} 
  \label{general_diagram}
\end{figure}

To extend the results for a generalised interconnected system, we consider $N$ interconnected switched systems. The schematics of the generalised interconnected system is shown in Fig. \ref{general_diagram}
The dynamics of $j^{th}$ subsystem in interconnection is given by,
\begin{equation}\label{general_pws_1}
\Sigma_j: \dot{\xi_j}=f_{ji}(\xi_j, u_{int}, u_{ext}, t), \hspace{0.2 cm}\xi_j(t)\in S_{ji} \subseteq{\mathbb{R}^{n}},
\end{equation}
where $j\in\{1, 2, 3,...N\}$ refers to the counter for a number of interconnected systems and $i \in\{1, 2\}$ refers to the counter for a number of modes in one subsystem. $S_{ji}$ is the region in state space corresponding to $j^{th}$ system $\Sigma_j$ and $i^{th}$ mode. Without loss of generality, $u_{ext}=\{u_1, u_2,...,u_N\}$ such that $u_j$ is the external input acting upon $\Sigma_j$, and $u_{int}$ are the output of other systems $\Sigma_l$ where $j\neq l$ in the interconnection which affect $\Sigma_j$. In this case, $x_l$, $l\neq j$ act like $u_{int}$. All other standard properties, such as the existence of the solution and smoothness of each mode, remain the same as that of \eqref{pws_1} for each $j$. The $\delta-$ ISS of the generalised interconnection is stated in the following corollaries.
\begin{corollary}
Consider $\Sigma_j$ satisfies all conditions stated in \eqref{interconneted_conditions} of Theorem \ref{theorem_3} for each $j\in\{1, 2,...N\}$, 
Then, \eqref{general_pws_1} is ISS with respect to $(\delta\xi_l, \delta u_j)$ with gain $\gamma_{jl}$ from $\delta\xi_l$ to $\delta\xi_j$, 
 for $l\neq j$, if there exist functions $\beta_j$ of class $\mathcal{KL}$ and $\gamma_{jl}, \gamma_j$ of class $\mathcal{K}_\infty$ i.e.,
\begin{equation}\label{ISS_general}   ||\delta\xi_j(t)||\leq\beta_j(||\delta\xi_j(0)||, t)+ \sum_{l=1}^{N}\gamma_{jl}(||\delta\xi_l||_t)+\gamma_j(||\delta u_j||_t),
\end{equation}
for all $t\geq =0$. The functions $\gamma_{jl}$ and $\gamma_{j}$ are called gains that characterise the effect of internal and external inputs, respectively. 
\end{corollary}
For vector-valued functions $\delta\xi=(\delta\xi_1^T, \delta\xi_1^T,..., \delta\xi_N^T)^T$, we define
\begin{equation}
    \|\delta\xi\|=\begin{bmatrix}
        \|\delta \xi_1\| \\.\\.\\\|\delta \xi_N\|  
    \end{bmatrix}, \hspace{0.2 cm} \gamma(||\delta u_{ext}||_t)=\begin{bmatrix}
\gamma_1(||\delta u||_t) \\.\\.\\\gamma_N(||\delta u||_t)  
    \end{bmatrix}, \hspace{0.2 cm}\textit{and}  \hspace{0.2 cm} \beta(s, t)=\begin{bmatrix}
\beta_1(s_1, t) \\.\\.\\\beta_N(s_N, t)  
    \end{bmatrix}
\end{equation}
The gain matrix $\Lambda^{ISS}:\mathbb{R}_+^N\mapsto \mathbb{R}_+^N$ is a map defined by
\begin{equation}\label{gain_matrix}
   \Lambda^{ISS}((s_1, s_2,..,s_N)^T):=\big(\sum_{l=1}^{N}\gamma_{1l}(s_l),...,\sum_{l=1}^{N}\gamma_{Nl}(s_l)\big)^T.
\end{equation}
Then, the ISS relationship in \eqref{ISS_general} can be rewritten in vectorised form as, 
\begin{equation}\label{ISS_general_new}   ||\delta\xi(t)||\leq\beta(||\delta\xi(0)||, t)+ \Lambda^{ISS}(||\delta\xi||_t)+\gamma(||\delta u_{ext}||_t).
\end{equation}
We introduce the following diagonal operator $D:\mathbb{R}_+^N\mapsto \mathbb{R}_+^N$ such that for $\rho_j\in$ $\mathcal{K}_\infty$, 
$j=\{1,2,..N\}$,
\begin{equation}\label{operator_d}
    D((s_1, s_2,..,s_N)^T)=\begin{bmatrix}
        (id+\rho_1)(s_1)\\.\\.\\(id+\rho_N)(s_N)
    \end{bmatrix}.
\end{equation}
We say that the gain matrix $\Lambda^{ISS}$  defined in \eqref{gain_matrix} satisfies the small gain condition \cite{ISS_small_gain_general, lyu2022small}, if there exists an operator $D$ as in \eqref{operator_d}, such that for all $s$, $s\neq 0$ we have
\begin{equation}\label{small_gain_condition_general}
    \Lambda^{ISS}\circ D(s):= \Lambda^{ISS}(D(s))\ngtr s.
\end{equation}
\begin{corollary}
Consider the general interconnected system \eqref{general_pws_1}, and suppose that each subsystem is $\delta$-ISS, i.e., condition \eqref{ISS_general} holds for all $j=1, 2,...N$ and the small gain condition stated in \eqref{small_gain_condition_general} is satisfied; then the generalised interconnected system \eqref{general_pws_1} is $\delta-$ISS.
\end{corollary}
The proof of results for the generalised case is standard and can be found in existing works such as \cite{ISS_small_gain_general}.
% where $\Hat{\beta}(t)$ is exponentially decaying, which proves the incremental stability of the overall system. For $\gamma_1\gamma_2<1,$ we need to ensure $\gamma_1\gamma_2=\frac{||F_2||}{|\mu(F_1)|}\frac{||G_2||}{|\mu(G_1)|}<1$. Recall the expression of $F_2$ and $G_2$ in \eqref{var_5_1} and \eqref{var_5_2}. If the interconnection matrices are bounded and satisfy the following relations,
% \begin{equation}
%     \Bigg\Vert\frac{\partial f_{11}(\xi_1, \xi_2, t)}{\partial \xi_2}\Bigg\Vert\leq \Tilde{k_1}, \hspace{0.2 cm}  \Bigg\Vert\frac{\partial f_{12}(\xi_1, \xi_2, t)}{\partial \xi_2}\Bigg\Vert\leq \Tilde{k_2},
% \end{equation}
% \begin{equation}
%     \Bigg\Vert\frac{\partial f_{21}(\xi_1, \xi_2, t)}{\partial \xi_1}\Bigg\Vert\leq \Tilde{m_1}, \hspace{0.2 cm}  \Bigg\Vert\frac{\partial f_{22}(\xi_1, \xi_2, t)}{\partial \xi_1}\Bigg\Vert\leq \Tilde{m_2}.
% \end{equation}
% then the small gain condition reduces to $$\frac{\Tilde{k}\Tilde{m}}{|\mu(F_1)||\mu(G_1)|}<1$$ where, $\Tilde{k}=max(\Tilde{k_1}, \Tilde{k_2})$ and $\Tilde{m}=max(\Tilde{m_1}, \Tilde{m_2}).  
\section{Results and Discussion}\label{Results}
Numerical examples are considered to verify the proposed results on the incremental stability of two switched nonlinear systems in interconnection. Numerous benchmark problems could be modelled in this framework, such as the interconnection of switching converters, cascaded bridge inverters, networked control systems, etc. The simulation study also considers the closed-form expression of the nonlinear switched system while explicitly considering the external input.
\subsection{Feedback Connection}
 Here, we consider a simple example of having two systems defined in \eqref{interconnected_feedback}, switching between two modes based on \textit{state-dependent} conditions. This numerical example is representative of complex switching systems connected in feedback.
\begin{equation}\label{example_2}
 \mathcal{G}:
    \begin{cases}
       \dot{\xi_1}=-4\xi_1-3\xi_1^2+2\xi_2+u_1 & \text{for $i=1$, when $\xi_1>0$},\\
    \dot{\xi_1}=-4\xi_1+3\xi_1^2+2\xi_2+u_1 & \text{for $i=2$, when $\xi_1<0$}.\\
    \end{cases}   
\end{equation}
\begin{equation}\label{example_4}
  \mathcal{H} : 
    \begin{cases}
       \dot{\xi_2}=-8\xi_2-3\xi_2^2+3\xi_1+u_2 & \text{for $i=1$, when $\xi_2>0$},\\
    \dot{\xi_2}=-8\xi_2+3\xi_2^2+3\xi_1+u_2 & \text{for $i=2$, when $\xi_2<0$},\\
    \end{cases}   
\end{equation}
where $u_1=0.3 \sin(10t)$ and $u_2=0.4\sin(10t)$ are sinusoidal input signals to the $\mathcal{G}$ and $\mathcal{H}$ respectively. For the $\mathcal{G}$-system in \eqref{example_2}, $H(\xi_1)=\xi_1=0$ is the switching surface. Here, $f_{11}=-4\xi_1-3\xi_1^2+4\xi_2+0.3sin(10t)$, $f_{12}=-4\xi_1+3\xi_1^2+4\xi_2+0.3sin(10t)$. Next, verifying the matrix measure condition as presented in theorem \ref{theorem_3}, $\frac{\partial f_{11}}{\partial \xi_1}=-4-6\xi_1$. When $\xi_1>0$, $\mu(\frac{\partial f_{11}}{\partial \xi_1})$ is negative. $\frac{\partial f_{12}}{\partial \xi_1}=-4+6\xi_1$. For the region $\xi_1<0$, $\mu(\frac{\partial f_{12}}{\partial \xi_1})$ is also negative. On the switching surface $\xi_1=0$, $\mu[(f_{11}-f_{12})\nabla H]$=$-6\xi_1^2=0$.

Similarly, for the $\mathcal{H}$-system, $H(\xi_2)=\xi_2=0$ is the switching surface. Here, $f_{21}=-8\xi_2-3\xi_2^2+4\xi_1+0.3\sin(10t)$, $f_{22}=-8\xi_2+3\xi_2^2+4\xi_1+0.3\sin(10t)$. Further, verifying the matrix measure conditions, $\frac{\partial f_{21}}{\partial \xi_2}=-8-6\xi_2$. When $\xi_2>0$, $\mu(\frac{\partial f_{21}}{\partial \xi_2})$ is negative. $\frac{\partial f_{22}}{\partial \xi_2}=-8+6\xi_2$. For the region $\xi_2<0$, $\mu(\frac{\partial f_{22}}{\partial \xi_2})$ is also negative. On the switching surface $\xi_2=0$, $\mu[(f_{21}-f_{22})\nabla H]$=$-6\xi_2^2=0$. Therefore, the conditions in Theorem \ref{theorem_3} are satisfied, which suffices the ISS of both $\mathcal{G}$ and $\mathcal{H}$ systems with respect to $(\delta\xi_2, \delta u_1)$ and $(\delta\xi_1, \delta u_2)$, respectively. This is equivalent to concluding $\delta$-ISS of $\mathcal{G}$ and $\mathcal{H}$ systems with respect to $(\xi_2, u_1)$ and $(\xi_1, u_2)$, respectively.

\begin{figure}[!ht]
		\centering
		\includegraphics[width=8 cm, height=3.9 cm]{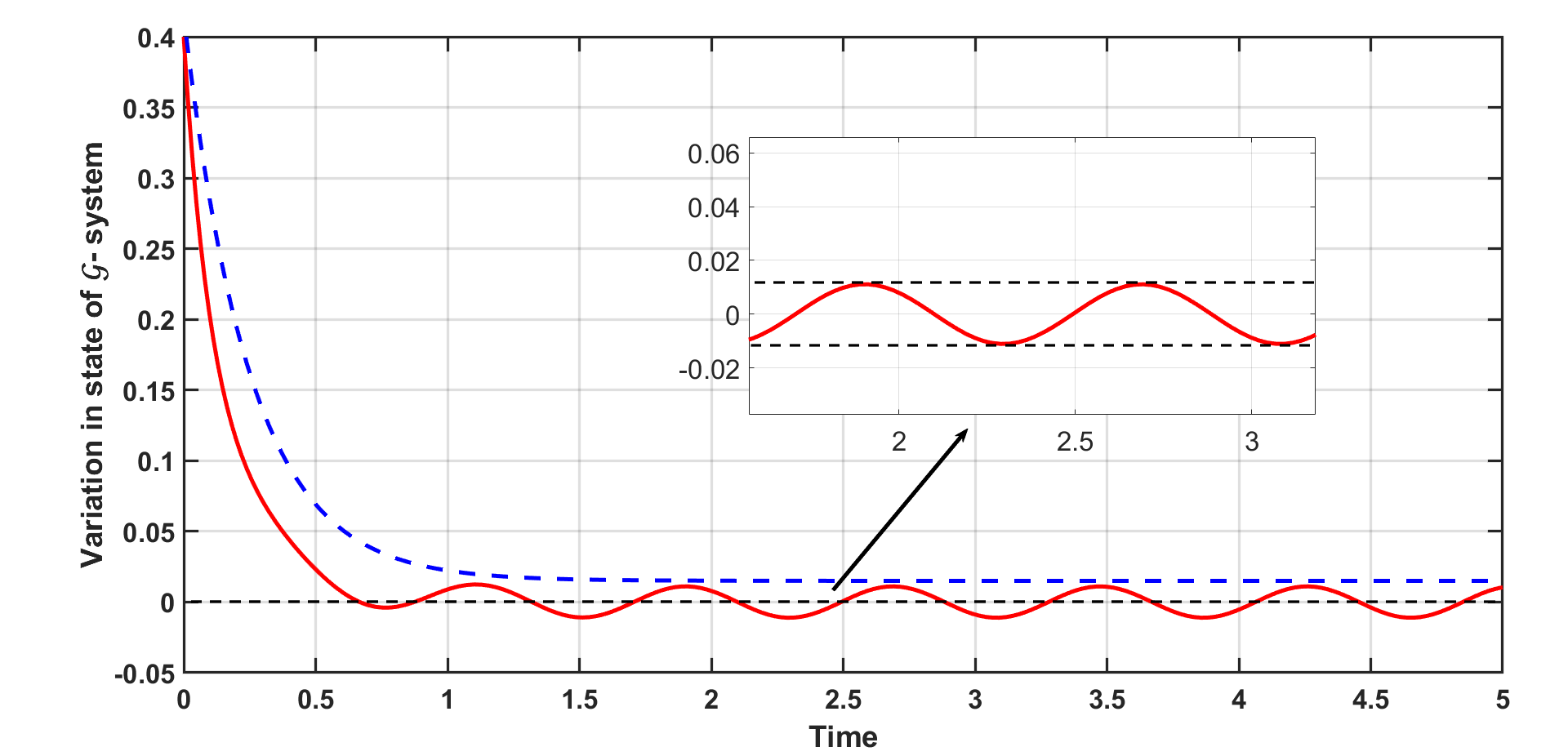}
		\caption{The time evolution of variational state $\delta \xi_1$ for feedback interconnection(Solid Curve).}
		\label{cont_1}%\vspace{-0.4cm}
\end{figure}

\begin{figure}[!ht]
		\centering
		\includegraphics[width=8 cm, height=3.9 cm]{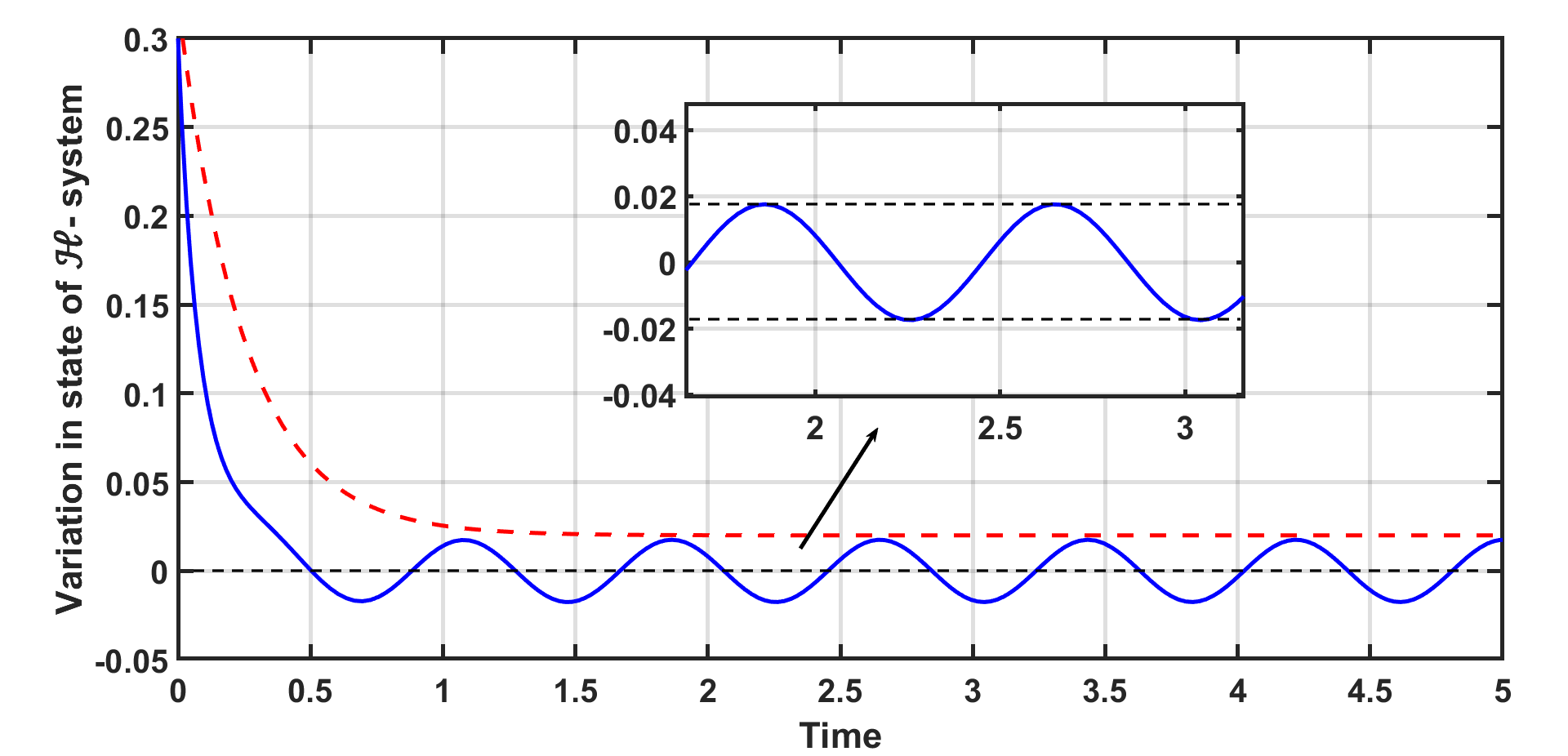}
		\caption{The time evolution of variational state $\delta \xi_2$ for feedback interconnection(Solid Curve).}
		\label{cont_2}%\vspace{-0.4cm}
\end{figure}

Next, we comment on the $\delta$-ISS of the overall interconnected system using the small-gain like condition in Theorem \ref{final_theorem}. First, we evaluate the interconnection terms such as $\frac{\partial f_{11}}{\partial \xi_2}=2$ and $\frac{\partial f_{12}}{\partial \xi_2}=2$ that are bounded. Similarly, $\frac{\partial f_{21}}{\partial \xi_1}=3$ and $\frac{\partial f_{22}}{\partial \xi_1}=3$. To satisfy the small gain condition in theorem \ref{final_theorem}, $\gamma_1$ and $\gamma_3$ for this numerical example are evaluated using the property $P2$ of matrix measure as given in \eqref{property_mu}. As this is a scalar system, $||F_2||=2$, $||G_2||=3$, and the absolute values of the matrix measure $|\mu(F_1)|=4$, $|\mu(G_1)|=8$. The gains are evaluated as $\gamma_1=0.5$ and $\gamma_3=0.375$. Hence, the small-gain condition $\gamma_1\gamma_2=0.1875<1$ is satisfied as in \ref{final_theorem}. To observe the evolution of the variational state corresponding to systems \eqref{example_2} and \eqref{example_4}, we allow incremental changes in input $u_1$ and $u_2$, respectively. From Fig.~\ref{cont_1} and Fig.~\ref{cont_2}, it can be observed that the variational states $\delta \xi_1(t)$ and $\delta \xi_2(t)$ corresponding to both $\mathcal{G}$ and $\mathcal{H}$ systems exponentially converge to a bound. The dotted curves in Fig.~\ref{cont_1} and Fig.~\ref{cont_2} represent the corresponding analytical bound as in \eqref{gronwal_4},\eqref{gronwal_5}. The bound depends on the values of the variational input $||\delta u_1(\tau)||_t$ and $||\delta u_2(\tau)||_t$. Therefore, the overall switched system in feedback interconnection is $\delta$-ISS.

\subsection{Cascade Connection}
Consider the special case of a generalised feedback interconnection, i.e. cascaded interconnection, which is representative of many applications. Only the simulation results are presented, as the theoretical results can be directly derived from the feedback connection. The example considered is
\begin{equation}\label{example_11}
  \mathcal{G} :
    \begin{cases}
       \dot{\xi_1}=-4\xi_1-9\xi_1^2+u_1 & \text{for $i=1$, when $\xi_1>0$},\\
      \dot{\xi_1}=-4\xi_1+9\xi_1^2+u_1 & \text{for $i=2$, when $\xi_1<0$}.\\
    \end{cases}   
\end{equation}
\begin{equation}
  \mathcal{H} : 
    \begin{cases}
       \dot{\xi_2}=-8\xi_2-3\xi_2^2+4\xi_1+u_2 & \text{for $i=1$, when $\xi_2>0$},\\
    \dot{\xi_2}=-8\xi_2+3\xi_2^2+4\xi_1+u_2 & \text{for $i=2$, when $\xi_2<0$},\\
    \end{cases}   
\end{equation}
where $u_1=0.3\sin(10t)$ and $u_2=0.4\sin(10t)$ are sinusoidal input signals to the $\mathcal{G}$ and $\mathcal{H}$ respectively. For the $\mathcal{G}$-system in the interconnection, $H(\xi_1)=\xi_1=0$ is the switching surface. Here, $f_{11}=-4\xi_1-9\xi_1^2+0.3\sin(10t)$, $f_{12}=-4\xi_1+9\xi_1^2+0.4\sin(10t)$. Next, verifying the matrix measure condition, we get $\frac{\partial f_{11}}{\partial \xi_1}=-4-18\xi_1$. When $\xi_1>0$, $\mu(\frac{\partial f_{11}}{\partial \xi_1})$ is negative. $\frac{\partial f_{12}}{\partial \xi_1}=-4+18\xi_1$. For the region $\xi_1<0$, $\mu(\frac{\partial f_{12}}{\partial \xi_1})$ is also negative. On the switching surface $\xi_1=0$, $\mu[(f_{11}-f_{12})\nabla H]$=$-18\xi_1^2=0$.

Similarly, for the $\mathcal{H}$-system, $H(\xi_2)=\xi_2=0$ is the switching surface. Here, $f_{21}=-8\xi_2-3\xi_2^2+4\xi_1$, $f_{22}=-8\xi_2+3\xi_2^2+4\xi_1$. Verifying the matrix measure conditions, $\frac{\partial f_{21}}{\partial \xi_2}=-8-6\xi_2$. When $\xi_2>0$, $\mu(\frac{\partial f_{21}}{\partial \xi_2})$ is negative. $\frac{\partial f_{22}}{\partial \xi_2}=-8+6\xi_2$. For the region $\xi_2<0$, $\mu(\frac{\partial f_{22}}{\partial \xi_2})$ is also negative. On surface $\xi_2=0$, $\mu[(f_{21}-f_{22})\nabla H]$=$-6\xi_2^2=0$.
Evaluating the Jacobian terms, $\frac{\partial f_{21}}{\partial \xi_1}=4$ and $\frac{\partial f_{22}}{\partial \xi_1}=4$ and using an argument similar to that in the feedback case, we can ensure Theorem \ref{theorem_3} and Theorem \ref{final_theorem} are satisfied.

\begin{figure}[!ht]
		\centering
		\includegraphics[width=8 cm, height=3.9 cm]{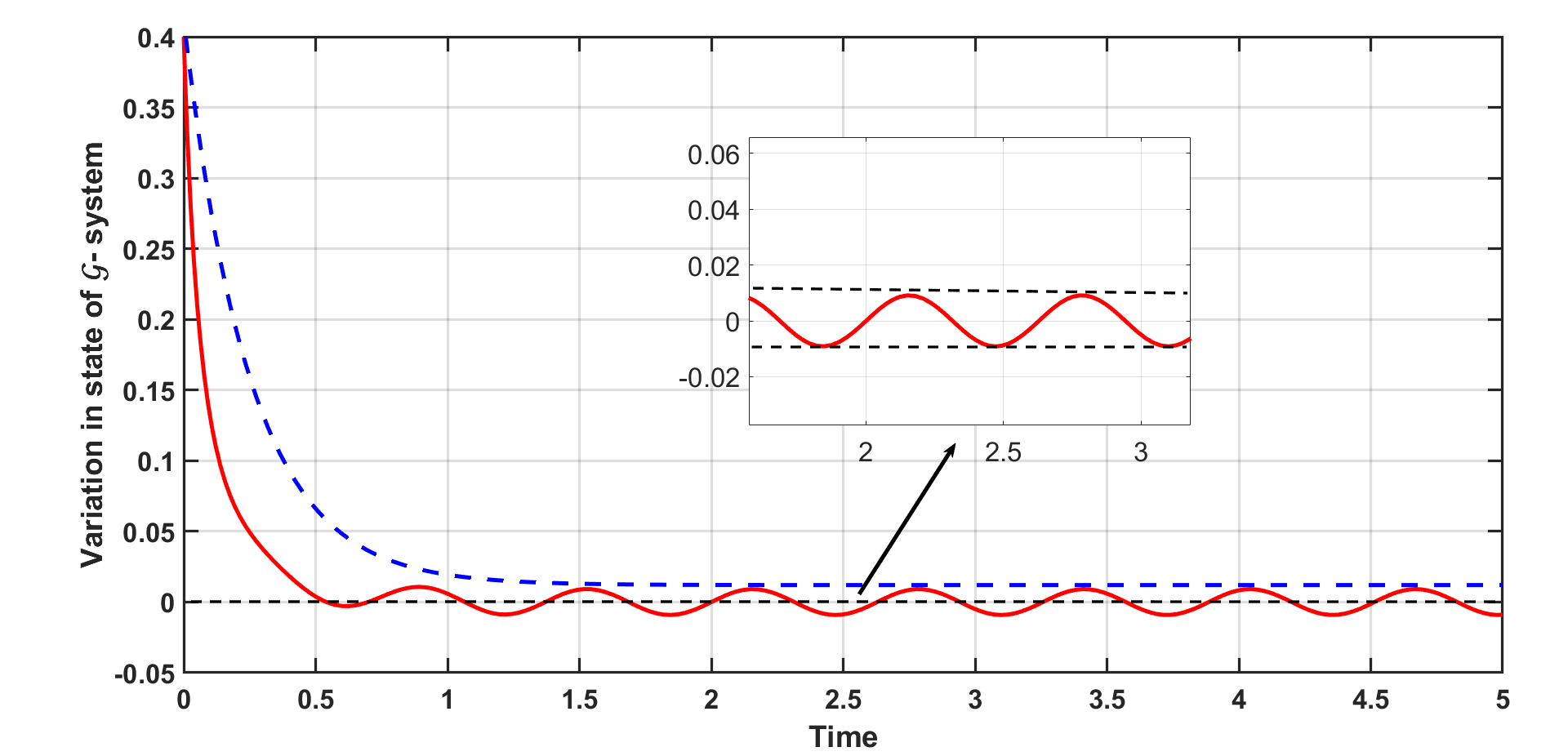}
		\caption{Time evolution of variational state $\delta \xi_1$ for cascade interconnection(Solid Curve)}.
		\label{cont_3}
\end{figure}

\begin{figure}[!ht]
		\centering
		\includegraphics[width=8 cm, height=3.9 cm]{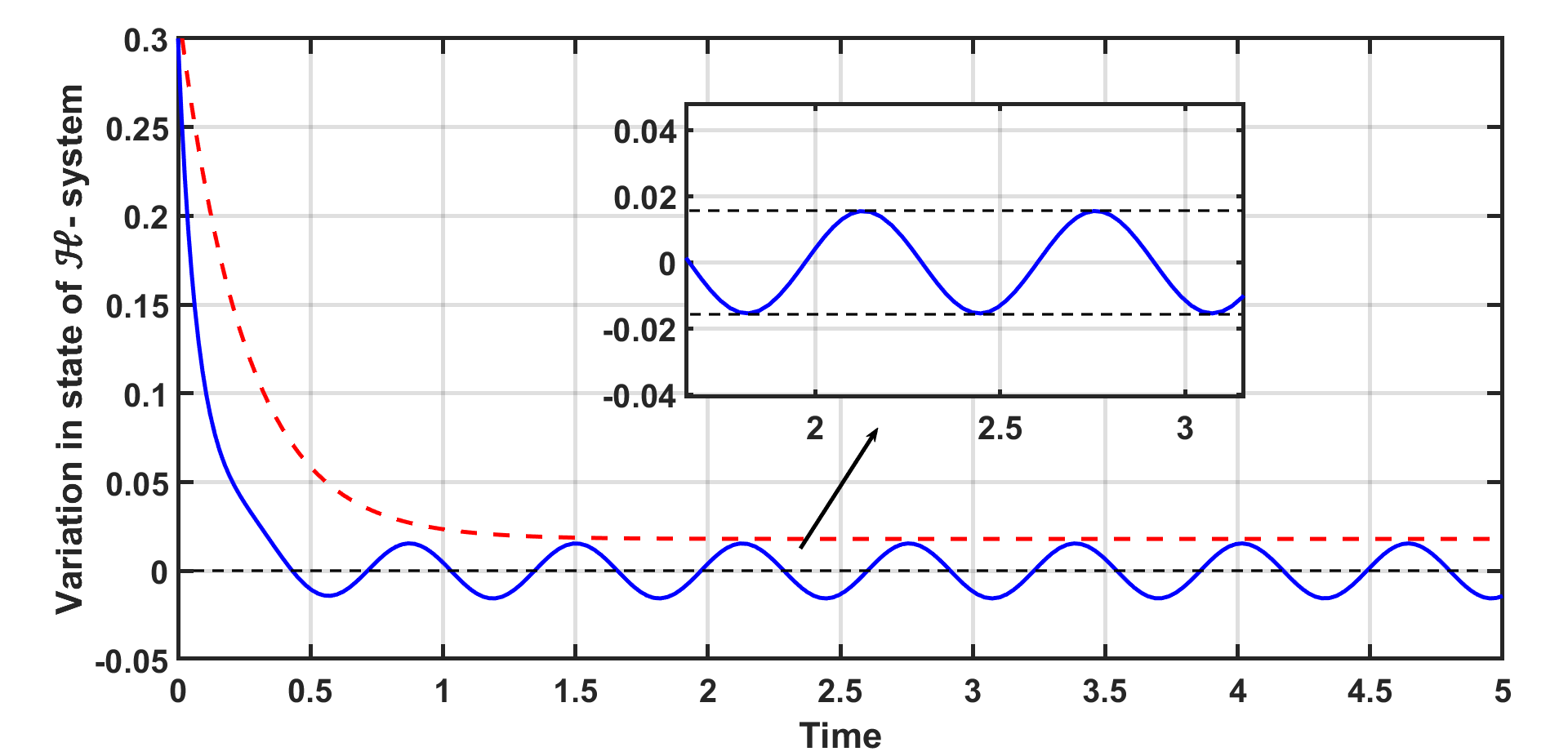}
		\caption{Time evolution of variational state $\delta \xi_2$ for cascade interconnection(Solid Curve).}
		\label{cont_4}
\end{figure}

The small-gain condition in cascade connection is automatically satisfied as it is a special case of feedback. In Fig.~\ref{cont_3} and Fig.~\ref{cont_4}, variational states $\delta \xi_1(t)$ and $\delta \xi_2(t)$ corresponding to both systems in cascade; the exponential convergence to a bound can be observed from the dotted curves. This signifies $\delta$-ISS of the overall cascaded switched nonlinear system.\\
%\vspace{-0.5 cm}

\begin{figure}[!ht]
		\centering
		\includegraphics[width=8 cm, height=3.9 cm]{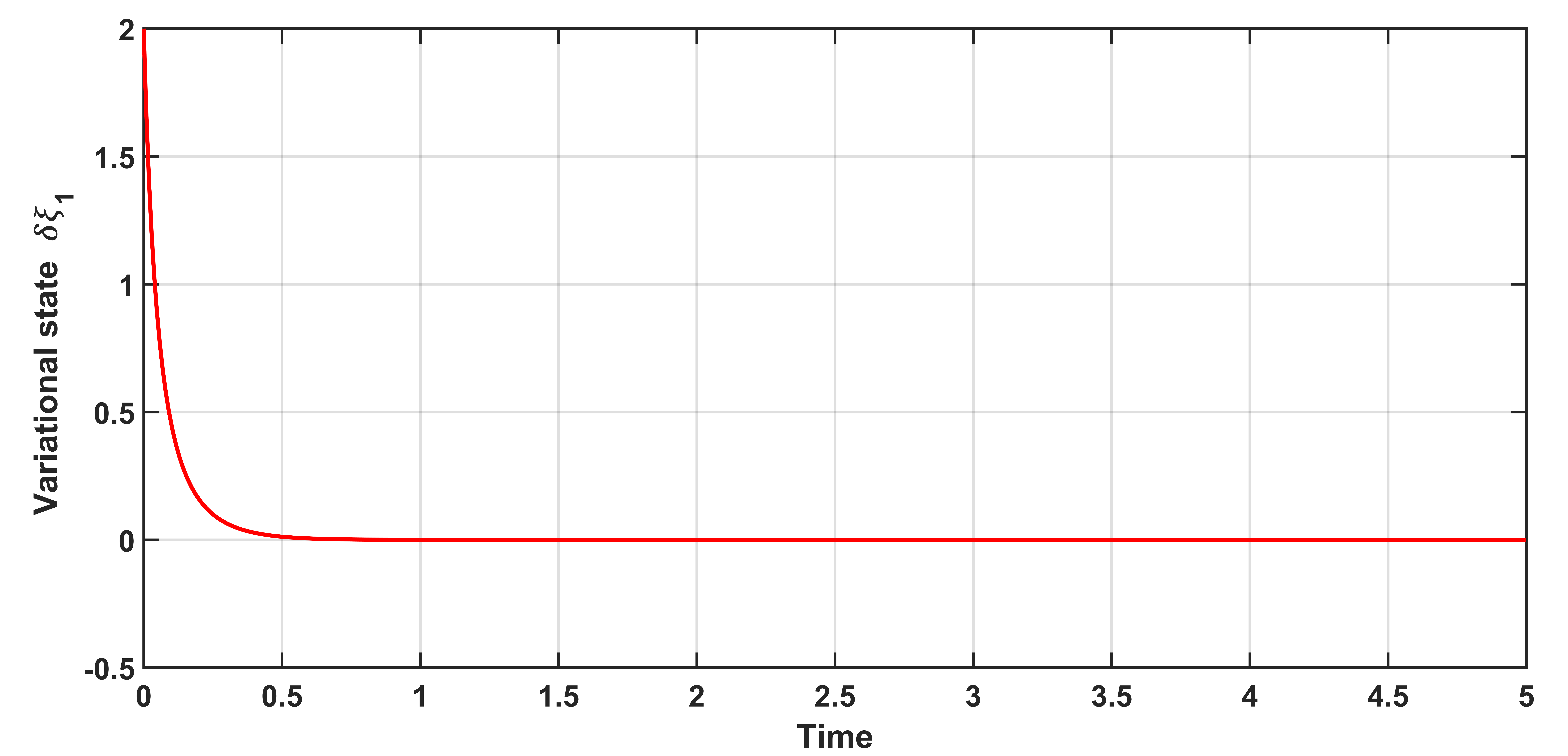}
		\caption{Evolution of variational state $\delta \xi_1$ when one of the modes is not contracting.}
		\label{cont_7}
\end{figure}

Throughout the paper, we assumed that the subsystems in a switched nonlinear system are individually contracting. Investigating whether one is not contracting and the other is contracting is intriguing. The idea is to derive a sufficient condition on a similar line as that in \cite{fiore2016contraction} for the switched system with mixed modes (i.e., contracting and noncontracting). We formulate this idea using the following remark, supported by a simulation study.\\
\begin{remark}\label{remark_special_case}
    Consider a special case of the variational system in \eqref{var_5_1}, which is not affected by the variational input ($\delta u_1$) or internal state ($\delta \xi_2$). Hence, no interconnection or external input is involved, causing no terms $F_2$ and $F_4$ in \eqref{var_5_1}. To prove the exponential stability of the variational system, following the third condition in \ref{theorem_3}, the matrix measure of term $F_3$ is zero. However, the first and second condition in the theorem demands the individual modes to be contractive ($\mu<0$). However, this can be relaxed if the sufficient condition in terms of the matrix measure condition stated in \eqref{relaxed} holds true.
    \begin{equation}\label{relaxed}
       \mu\Bigg( \psi(\xi_1)\frac{\partial f_{11}}{\partial \xi_1}+\Gamma(\xi_1)\frac{\partial f_{22}}{\partial \xi_1}\Bigg)<0.
    \end{equation}
Even if one of the modes is not contracting, satisfying the stated inequality ensures that the variational state will exponentially converge to zero. Hence, we may ensure the contractivity of a bimodal switched system. 
Consider an example of a switched system
\begin{equation}\label{example_1}
    \begin{cases}
       \dot{\xi_1}=-8\xi_1-3\xi_1^2 & \text{for $i=1$, when $\xi_1>0$},\\    
       \dot{\xi_1}=4\xi_1 & \text{for $i=2$, when $\xi_1<0$}.\\
    \end{cases}   
\end{equation}
The vector fields in the example in \eqref{example_1} are $f_{11}=-8\xi_1-6\xi_1^2$ and $f_{12}=4\xi_1$. Evaluating the Jacobian $\frac{\partial f_{11}}{\partial \xi_1}=-8-3\xi_1$. For $\xi_1>0$, $\mu(\frac{\partial f_{11}}{\partial \xi_1})<0$, which satisfies the sufficient condition in Theorem \ref{theorem_3}. For the Jacobian $\frac{\partial f_{12}}{\partial \xi_1}=4$, the matrix measure $\mu(\frac{\partial f_{12}}{\partial \xi_1})>0$, which violates Theorem \ref{theorem_3}. However, if we observe the matrix measure of Jacobians, there exist some $\psi(\xi_1)$ and $\Gamma(\xi_1)$ such that the relation in \eqref{relaxed} gets satisfied, which suffices for the contraction of the switched system. The evolution of the variational state $\delta \xi_1$ is shown in Fig.~\ref{cont_7}, which also shows exponential convergence.
\end{remark}

\textbf{Key Highlights from the Simulation Study}: The simulation study has two benchmark problems where we could show the efficacy of the proposed work for feedback and cascade interconnection. Unlike \cite{van2023small}, which considers a linear system in feedback with a hybrid integrator system, constructs an overall switched interconnected system, the class of system we consider is more general and complex and can be representative of many applications. In that scenario, by using the small gain condition derived, we could achieve exponential convergence of the variational state($\delta$-ISS) as can be seen from Fig.~\ref{cont_1}, Fig.~\ref{cont_2}, Fig.~\ref{cont_3} and Fig.~\ref{cont_4}.  The analytical bounds(red dashed curves in each Figure) represent the exponentially decaying envelope, which are quite tight, reflecting the effectiveness of the proposed results. In contrast, most of the existing works related to interconnected systems talk about asymptotic convergence. Finally, we explore a special case where there are mixed contracting and noncontracting modes in a single nonlinear switched system, where we could show exponential convergence in Fig.~\ref{cont_7}, provided a new sufficient condition is satisfied, a contrasting result compared to \cite{fiore2016contraction}.
\section{CONCLUSION}
The results proposed in this paper exploited the notion of contraction theory to study the incremental stability of interconnected nonlinear switched systems. We considered individual subsystems to be bimodal switched systems, and the nature of the switched considered is state-dependent switching. Analytical sufficient conditions are derived in order to ascertain the incremental input-to-state stability ($\delta$-ISS) of interconnected switched nonlinear systems in the presence of external inputs. Variational dynamics using Contraction theory have been used to obtain the set of mathematical conditions which ensure exponential convergence, whereas most of the existing literature related to $\delta-$ISS of interconnected systems comment on asymptotic convergence. To propose the results, different interconnections are considered that are used extensively in different applications. In the case of feedback interconnection of bimodal switched systems, we propose a small gain condition to comment on the $\delta$-ISS of the overall system. To add generality, the results are extended to general interconnections, which can be representative of large-scale systems, and corresponding small-gain conditions are stated. This explains how incremental stability is useful to enhance the scalability of a system under consideration. In addition, we extend the results for a class of cascaded systems. As an interesting extension, a special case in which one of the modes need not be contracting, very unlikely in the literature, is introduced with a supporting example.

Extending the results for the general higher-order interconnections of switched systems compared to existing results is under investigation. Our approach in this paper has a limitation as the analysis relies on exact model knowledge, which is not available in many applications. In the future, we plan to use data-driven techniques such as the Gaussian process \cite{sundarsingh2024backstepping} or neural network-based approaches \cite{basu2025formally_verification,basu2025formally,basu2025neural} to comment on $\delta-$ISS of an overall interconnected system. It is also planned to consider time-dependent switched interconnected systems in future.
\section*{Declarations}

\textbf{Conflict of interest :}
The authors declare that they have shared no conflict of interest.

\textbf{Funding :}
No funding has been received to assist with this research work.

\textbf{Data Availability :}
All the data relevant to this paper’s findings are provided within the document itself.                            % on the last page of the document manually. It shortens

%%%%%%%%%%%%%%%%%%%%%%%%%%%%%%%%%%%%%%%%%%%%%%%%%%%%%%%%%%%%%%%%%%%%%%%%%%%%%%%%

%%%%%%%%%%%%%%%%%%%%%%%%%%%%%%%%%%%%%%%%%%%%%%%%%%%%%%%%%%%%%%%%%%%%%%%%%%%%%%%%

%%%%%%%%%%%%%%%%%%%%%%%%%%%%%%%%%%%%%%%%%%%%%%%%%%%%%%%%%%%%%%%%%%%%%%%%%%%%%%%%

%  \section{ACKNOWLEDGMENT}

% The authors thank the members of the Control and Automation group of IIT Delhi for continuous motivation and discussion.

%%%%%%%%%%%%%%%%%%%%%%%%%%%%%%%%%%%%%%%%%%%%%%%%%%%%%%%%%%%%%%%%%%%%%%%%%%%%%%%%

 \bibliographystyle{IEEEtran}
 \bibliography{reference} % Entries are in the refs.bib file
\end{document}